\documentstyle[12pt,psfig]{article}
\newcommand{\beq}{\begin{equation}}
\newcommand{\eeq}{\end{equation}}
\newcommand{\beqa}{\begin{eqnarray}}
\newcommand{\eeqa}{\end{eqnarray}}
\newcommand{\etal}{{\it et al.}}

\def\lsim{\:\raisebox{-0.5ex}{$\stackrel{\textstyle<}{\sim}$}\:}
\def\gsim{\:\raisebox{-0.5ex}{$\stackrel{\textstyle>}{\sim}$}\:}

\begin{document}
\begin{flushright}
 SINP/TNP/01-22
\end{flushright}
\bigskip
\begin{center}
{\bf Three Generation Neutrino Oscillation Parameters after  SNO}
\\[2cm] Abhijit Bandyopadhyay\footnote{abhi@theory.saha.ernet.in},
Sandhya Choubey \footnote{sandhya@theory.saha.ernet.in},
Srubabati Goswami \footnote{sruba@theory.saha.ernet.in},
Kamales Kar \footnote{kamales@theory.saha.ernet.in}
\\[1cm]  Saha Institute of Nuclear Physics, \\ 1/AF,
Bidhannagar, Kolkata 700 064, INDIA
\end{center}

\vspace{1cm}

\begin{center}
{\bf Abstract}
\end{center}
\bigskip

We examine the solar neutrino problem in the context of the
realistic three neutrino mixing scenario including the SNO charged
current (CC) rate. The two independent mass squared differences
$\Delta m^2_{21}$ and $\Delta m^2_{31} \approx \Delta m^2_{32}$
are taken to be in the solar and atmospheric ranges respectively.
We incorporate the constraints on $\Delta$m$^2_{31}$ as obtained
by the SuperKamiokande atmospheric neutrino data and determine the
allowed values of $\Delta m^2_{21}$, $\theta_{12}$ and
$\theta_{13}$ from a combined analysis of solar and CHOOZ data.
Our aim is to probe the changes in the values of the mass and
mixing parameters with the inclusion of the SNO data as well as
the changes in the two-generation parameter region obtained from
the solar neutrino analysis with the inclusion of the third
generation. We find that the inclusion of the SNO CC rate  in the
combined solar + CHOOZ analysis puts a more restrictive bound on
$\theta_{13}$. Since the allowed values of $\theta_{13}$ are
constrained to very small values by the CHOOZ experiment there is
no qualitative change over the two generation allowed regions in
the $\Delta m^2_{21} - \tan^2 \theta_{12}$ plane. The best-fit
comes in the LMA region and no allowed area is obtained in the SMA
region at 3$\sigma$ level from combined solar and CHOOZ analysis.\\

PACS numbers(s): 14.60.Pq, 12.15.Ff, 26.65.+t.

\section{Introduction}
The recent results on charged current measurement from Sudbury
Neutrino Observatory (SNO) \cite{sno}  have  confirmed the solar
neutrino shortfall as observed in the earlier  experiments
\cite{sksolar, cl, kam,ga}.  A comparison of the SuperKamiokande
and SNO results establishes the presence of non-electron flavor
component in the solar neutrino flux received at earth (at more
than  3$\sigma$ level)  in a model independent manner \cite{sno,
barger,lisisno, giunti}. Neutrino oscillation provides the most
popular explanation to this anomaly. Two generation analysis of
the solar neutrino data including the SNO results has been
performed by various groups \cite{lisisno, bgc, bcgk, strumia,seis,ks,eind2,
baysian}. All these analyses agree that the best
description to the data on the total rates and the day/night
spectrum data of the SuperKamiokande (SK) collaboration is
provided by the Large Mixing Angle (LMA) MSW solution ($\Delta
m^2_{\odot} \sim 10^{-5}$ eV$^2$), though the low
$\Delta$m$^2_\odot$ solution (LOW-QVO) ($\Delta m^2_{\odot} \sim
10^{-9} - 10^{-7} $ eV$^2$) and the vacuum oscillation (VO)
solutions ($\Delta m^2 \sim 4.5 \times 10^{-10}$ eV $^2$) are also
allowed. The Small Mixing Angle (SMA) MSW solution is largely
disfavoured with no allowed contour in the mass-mixing plane at
the $3 \sigma$ level \footnote{Only exception is the analysis of
\cite{bgc} which get a small allowed region for the SMA solution
due to a slight difference in the treatment of the data.}. On the
other hand, for the explanation of the atmospheric neutrino
anomaly the two generation oscillation analysis of the atmospheric
neutrino data requires $\Delta$m$^2_{atm} \sim$ 10 $^{-3}$ eV$^2$
\cite{skatm}. Since the allowed ranges of $\Delta$m$^2_\odot$ and
$\Delta$m$^2_{atm}$ are completely non-overlapping, to explain the
solar and atmospheric neutrino data simultaneously by neutrino
oscillation, one requires at least two independent mass-squared
differences and consequently three active neutrino flavors which
fits very nicely with the fact that to date we have observed three
neutrino flavours in nature. Thus to get the complete picture of
neutrino masses and mixing a three generation analysis is called
for. Apart from the solar and atmospheric neutrinos positive
evidence for neutrino oscillation is also published by the LSND
experiment \cite{lsnd} and although there had been several
attempts to explain all the three evidences in a three generation
picture it is now widely believed that to accommodate the LSND
results one has to introduce an additional sterile neutrino
\cite{sterile1,sterile2}. For the purpose of this analysis we ignore the
LSND results. We incorporate the negative results from the CHOOZ
reactor experiment on the measurement of $\bar{\nu_{e}}$
oscillation  by disappearance  technique \cite{chooz}. CHOOZ is sensitive to
$\Delta$m$^2_{CHOOZ} \gsim 10^{-3}$ eV$^2$ which is the range
probed in the atmospheric neutrino measurements and together they
can  put important constraints  on  the three neutrino mixing
parameters.
 We consider the three flavour picture with
\\ $\bullet~~ \Delta m^2_{21} = \Delta m^2_\odot$ , $\bullet~~
\Delta m^2_{31} = \Delta m^2_{CHOOZ} \simeq \Delta m^2_{atm} =
\Delta m^2_{32}$. \\

Three flavor oscillation analysis of
solar, atmospheric
and
CHOOZ data assuming this mass spectrum
was performed in pre SNO era by different groups
$\cite{fl3g,gg3g,presno3}$. We investigate the impact of  the charged
current measurement at SNO on neutrino mass and mixing in a three
flavor scenario and present the most up to date status of the
allowed values of three flavor oscillation parameters.

The plan of the paper is as follows. In section 1 we present the
relevant probabilities. In section 3 we discuss the
$\chi^2$-analysis method and the results. We end in section 4 with
some discussion and conclusions.

\section{Calculation of Probabilities}
The three-generation mixing matrix that we use is \beqa U & = &
R_{23}R_{13}R_{12} \nonumber\\ & = &\pmatrix {c_{13}c_{12} &
s_{12}c_{13} & s_{13} \cr -s_{12}c_{23} - s_{23} s_{13} c_{12} &
c_{23} c_{12} - s_{23} s_{13} s_{12} & s_{23} c_{13} \cr s_{23}
s_{12} - s_{13} c_{23} c_{12} & -s_{23} c_{12} - s_{13} s_{12}
c_{23} & c_{23} c_{13} \cr} \label{mix} \eeqa where we neglect the
CP violation phases. This is justified as one can show that the
survival probabilities $P_{ee}$ of the electron neutrinos do not
depend on these phases. The above choice has the advantage that
the matrix elements U$_{e1}$, U$_{e2}$ and U$_{e3}$ relevant for
the solar neutrino problem becomes independent of $\theta_{23}$
while the elements U$_{e3}$, U$_{\mu3}$ and U$_{\tau3}$ relevant
for the atmospheric neutrino problem are independent of
$\theta_{12}$. The  mixing angle common to both solar and
atmospheric neutrino  sectors is $\theta_{13}$ which, as we will
see, is constrained severely by the CHOOZ data.

\subsection{Solar Neutrinos}

The general expression for the survival amplitude for an electron
neutrino arriving on the earth from the sun,  in presence of three
neutrino flavours is given by \cite{qvo}
\begin{equation}
A_{ee} = A_{e1}^\odot  A_{11}^{vac} A_{1e}^\oplus + A_{e2}^\odot
A_{22}^{vac} A_{2e}^\oplus  +  A_{e3}^\odot A_{33}^{vac}
A_{3e}^\oplus \label{amp}
\end{equation}
where $A_{ek}^\odot$ gives the probability amplitude of $\nu_e
\rightarrow \nu_k$ transition at the solar surface, $A_{kk}^{vac}$
gives the transition amplitude from the solar surface to the earth
surface,$A_{ke}^\oplus$ denotes the $\nu_k\rightarrow \nu_e$
transition amplitudes inside the earth. One can write the
transition amplitudes in the sun as an amplitude part times a
phase part
\begin{equation}
A_{ek}^\odot = a_{ek}^\odot e^{-i \phi^\odot_{k}}
\label{aeksun}
\end{equation}
${a_{ek}^\odot}^2$ can be expressed as
\begin{equation}
{a_{ek}^{\odot}}^2 = \sum_{j=1,2,3} X_{kj} {{U^\odot_{je}}}^2
\label{asun}
\end{equation}
where $X_{kj}$ denotes the non-adiabatic jump probability between
the j$^{th}$ and k$^{th}$ state and $U_{je}^\odot$ denotes the
mixing matrix element between the flavour state $\nu_e$ and the
mass state $\nu_j$ in sun. 
$A_{kk}^{vac}$ is given by
\begin{equation}
A_{kk}^{vac} = e^{-i E_{k} (L - R_{\odot})}
\end{equation}
where $E_k$ is the energy
of the state $\nu_k$, $L$ is the
distance between the center of the Sun and Earth and $R_\odot$ is
the solar radius.
For a two slab model of the earth --- a mantle and core with
constant densities of 4.5 and 11.5 gm cm$^{-3}$ respectively, the
expression for $A_{ke}^{\oplus}$ can be written as (assuming the
flavor states to be continuous across the boundaries)\cite{petearth},
\begin{eqnarray}
A_{ke}^{\oplus}&=& \sum_{\stackrel{i,j,l,}{\alpha,\beta,\sigma}}
U_{el}^M e^{-i\psi_l^M} U_{\alpha l}^M U_{\alpha i}^C
e^{-i\psi_i^C} U_{\beta i}^C U_{\beta j}^M
e^{-i\psi_j^M}
U_{\sigma j}^M U_{\sigma k} \label{aearth}
\end{eqnarray}
where ($i,j,l$) denotes mass eigenstates and
($\alpha,\beta,\sigma$) denotes flavor eigenstates, $U^M$ and
$U^C$ are the mixing matrices in the mantle and the core
respectively and $\psi^M$ and $\psi^C$ are the corresponding
phases picked up by the neutrinos as they travel in the mantle and
the core of the Earth.
\beqa
P_{ee} & = &  |A_{ee}|^2 \nonumber \\
       & = &\Sigma_{k}{a_{ei}^\odot |A_{ke}^\oplus|^2}
+ \sum_{l>k}{2 a_{ek}^\odot a_{el}^{\odot} Re[A_{ke}^{\oplus}
A_{le}^{\oplus} e^{i(E_l - E_k)(L - R_{\odot})}
e^{i(\phi_l^{\odot}- \phi_k^{\odot})}]}
\eeqa
This is the most general expression for the
probability \cite{stp}.
Since for our case $\Delta_{31}$ $\approx
\Delta_{32}$ is $\sim$ $10^{-3}$ eV$^2$ the phase terms $e^{i (E_3
- E_1)(L-R_{\odot})}$ and $e^{i( E_3 -E_2)(L - R_\odot)}$ average
out to zero. Therefore the probability simplifies to
\beqa
P_{ee} & = & a_{e1}^{\odot 2}|A_{1e}^{\oplus}|^2 +
a_{e2}^{\odot 2}|A_{2e}^{\oplus}|^2 +
a_{e3}^{\odot 2}|A_{3e}^{\oplus}|^2 \nonumber\\
 &  & + 2a_{e1}^{\odot} a_{e2}^{\odot} Re[A_{1e}^\oplus {A_{2e}^
\oplus}^{*} e^{i(E_2-E_1)(L-R_\odot)}
e^{i(\phi_{2}^\odot-\phi_{1}^\odot)}] \label{pr} \eeqa
The mixing matrix elements in matter 
are different from those in vacuum and it is in general a
difficult task to find the matter mixing angles and eigenvalues
for a $3 \times 3$ matrix. However  in our case since
$\Delta$m$_{31}^2
>>  \Delta$m$_{21}^2 \approx$ the matter potential in sun , the
$\nu_3$ state experiences almost no matter effect and  MSW
resonance can occur between $\nu_2$ and $\nu_1$ states. Under this
approximation the three generation survival probability for the electron 
neutrino  can be expressed as, 
 \beqa P_{ee} = c_{13}^4
P_{ee}^{2gen} + s_{13}^4 \eeqa
where $P_{ee}^{2gen}$ is of the two generation form in the mixing angle 
$\theta_{12}$.  
\beqa
P_{ee}^{2gen} = P_{ee}^{day} + \frac{(2 P_{ee}^{day} - 1)(\sin^2 \theta_{12} -
|A_{2e}^{\oplus}|^2)} {\cos{2\theta_{12}}},
\label{pee}
\eeqa
where 
\begin{equation}
P_{ee}^{day} = 0.5+[0.5
-\Theta(E - E_{A})X_{12}]{\cos2\theta_{12}^{\odot}}{\cos2\theta_{12}},
\label{pf}
\end{equation}
with
\beq \tan 2\theta_{12}^\odot = \frac{\Delta
m^2_{21} \sin 2\theta_{12}}{\Delta m^2_{21}\cos2\theta_{12} -A
c_{13}^2} \label{thetam} \eeq
where  $A$ denotes the matter potential, \beq A =
2\sqrt{2}G_{F}n_{e}^\odot E \eeq here $n_{e}^\odot$ is the
electron density in the sun, $E$ the neutrino energy, and $\Delta
m^2_{21}$ (= ${m_{2}^2 - m_{1}^2}$) the mass squared difference in
vacuum. 
The jump probability $X_{12}$ continues to be 
given by the two-generation expression and for this we use the analytic 
expression given in \cite{petcov}. 
$E_{A}$  in the Heaviside function $\Theta$ 
gives the minimum $\nu_e$ energy that can encounter a resonance
inside the sun and is given by
\begin{equation}
E_{A} = {\Delta{m^2_{21}}\cos2\theta_{12}}/{2\sqrt{2}G_{F}n_{e}|_{pr}},
\end{equation}
gives the minimum $\nu_e$ energy that can encounter a resonance
inside the sun, $n_{e}|_{pr}$ being the electron density at the
point of production. 
In the
limit $\theta_{13}$ = 0 one recovers the two generation limit.

\subsection{The Probability for CHOOZ}
The survival
probability relevant for the CHOOZ experiment for the three generation case 
is
\beqa 
P_{ee} & =  & 1 -
c_{13}^4 \left[\sin^2 2\theta_{12} \sin^2\frac{\Delta m_{21}^2 L}{4E}
\right] 
 - \sin^2 2\theta_{13} \sin^2 \frac{\Delta m_{31}^2 L}{4E} \nonumber \\
&  & +sin^2 2\theta_{13} s_{12}^2 \left[\sin^2 \frac{\Delta m_{31}^2 L}{4 E} 
 - \sin^2 \frac{(\Delta m^2_{31} - \Delta m^2_{21}) L}{4E} \right] 
\eeqa 
Since the
average energy of the neutrinos in the CHOOZ experiment is $\sim$
1 MeV and the distance traveled by the neutrinos is of the order
of 1 Km the $\sin^2 (\frac{\Delta m_{21}^2 L}{4E})$ term is important only for
$\Delta m^2_{21} \gsim 3 \times 10^{-4}$ eV$^2$. 
The last term in the above expression is an 
interference term  between both mass scales  \cite{piai} and  
is absent if one uses the approximation $\Delta_{31} = \Delta_{32}$
and is often ignored. 

\section{The $\chi^2$ analysis}

The definition of $\chi^2_{\odot}$ used in our fits is
\beqa
\chi^2_\odot & =  &
\sum_{i,j=1,4} (R_i^{\rm th} - R_i^{\rm exp})
{\left[{(\sigma^{rates}_{ij})}^2\right]}^{-1} (R_j^{\rm th} -
 R_j^{\rm exp})
 \nonumber \\
        &   & + \sum_{i,j=1.38} (X_n S_i^{\rm th} - S_i^{\rm exp})
{\left[{(\sigma^{spm}_{ij})}^2\right]}^{-1} (X_n S_j^{\rm th} - S_j^{\rm exp})
\label{two} \eeqa
where $R_i^\xi$ ($\xi$ = th or exp)
 denote the total rate
while $S_{i}^\xi$
denote the  SK spectrum in the i$^{th}$ bin.
Both the experimental and theoretical values of
the fitted quantities are normalised relative to the BPB00
\cite{bpb00} predictions.
The experimental values for the total
rates are the ones shown in Table 1, while the SK day-night
spectra are taken from \cite{sksolar}.  The error matrix
${(\sigma^{rates})^2}$ contains the experimental errors, the
theoretical errors (which includes error in the capture cross-sections and
the astrophysical uncertainties in BPB00 predictions) along with their
correlations. It is evaluated using the procedure of
\cite{flap}. The error matrix for the spectrum ${(\sigma^{spm})}^2$
contains the correlated and uncorrelated errors as discussed in
\cite{ggerrorsolar}. The details of the solar code used is described in
\cite{bcgk,sg,seis,eind2}.
We vary the normalisation of the SK spectrum $X_n$ as a
free parameter to avoid double counting with the SK data on total
rate.  Thus there are $(38 - 1)$ independent data points from the
SK day-night spectrum along with the 4 total rates giving a total
of 41 data points. For the analysis of only the solar data in
the three-generation scheme, we have $(41-3)$ degrees of freedom (DOF).
The best-fit values of parameters and the
$\chi^2_{min}$ are
\begin{itemize}
\item $\Delta m^2_{21} = 4.7 \times 10^{-5}$ eV$^2$,
$\tan^2{\theta_{12}}$ = 0.375,
$\tan^2{\theta_{13}}$ = 0.0, $\chi^2_{min}$ = 33.42
\end{itemize}
Hence the best-fit comes in the two-generation limit presented in
\cite{bcgk,seis,eind2}.

We next incorporate the results from the CHOOZ reactor experiment
\cite{chooz}.
The definition of $\chi^2_{CHOOZ}$ is given by \cite{chichooz}
\begin{equation}
\chi^2_{CHOOZ} = \sum_{j=1,15}(\frac{x_{j} - y_{j}}{\Delta
x_{j}})^2
\end{equation}
where $x_{j}$ are the experimental values, $y_{j}$ are the
corresponding theoretical predictions, $\Delta x_{j}$ are the
1$\sigma$ errors in the experimental quantities and the sum is
over 15 energy bins of data of the CHOOZ experiment \cite{chooz}.
The global $\chi^2$ for solar+CHOOZ analysis is defined as
\beq
\chi^2_{global} = \chi^2_{\odot} + \chi^2_{CHOOZ} \eeq
The total
number of data points for combined solar and CHOOZ analysis is
therefore 41+15 = 56. The solar+CHOOZ analysis depends on
$\Delta m_{21}^2$, $\Delta m_{31}^2$, $\theta_{12}$ and $\theta_{13}$.
For unconstrained $\Delta m^2_{31}$, the $\chi^2_{min}$ and the
best-fit values are
\begin{itemize}
\item $\Delta m^2_{21} = 4.7 \times 10^{-5}$ eV$^2$,
$\tan^2{\theta_{12}}$ = 0.374, $\Delta m^2_{31} = 1.35
 \times 10^{-3}$ eV$^2$,\\ $\tan^2{\theta_{13}} = 1.74 \times 10^{-3}$,
$\chi^2_{min} = 39.75$
\end{itemize}
However the atmospheric neutrino data imposes strong constraints on the
allowed range of $\Delta m_{31}^2$. The combined analysis of the
1289 day atmospheric data and the CHOOZ data restricts allowed
$\Delta m_{31}^2$ in the range $[1.5,6]\times 10^{-3}$ eV$^2$ at 99\% C.L.
\cite{fl3g}.
Thus the best-fit $\Delta m^2_{31} = 1.35 \times 10^{-3}$ that we obtain
from the solar+CHOOZ analysis falls outside the allowed range.
If we restrict the range of
$\Delta m_{31}^2$ from the combined analysis of the
atmospheric+CHOOZ analysis \cite{fl3g} then the $\chi^2_{min}$ and
the best-fit parameters obtained from the combined solar+CHOOZ
analysis are
\begin{itemize}
\item
$\Delta m_{21}^2=4.7 \times 10^{-5}$ eV$^2$,
$\tan^2{\theta_{12}}$ = 0.374, $\Delta m^2_{31} = 1.5 \times
10^{-3}$ eV$^2$,\\  $\tan^2{\theta_{13}} = 1.46 \times 10^{-3}$,
$\chi^2_{min} = 39.75$
\end{itemize}

Thus the best-fit for the solar+CHOOZ analysis
comes almost at the two generation limit, with the best-fit $\Delta m^2_{31}$
at the lower limit of the allowed range.
For 52 DOF this solution is allowed at 89.33\%.
The improvement
in the goodness of fit (GOF)
in comparison to the two flavour analysis presented in
\cite{bcgk,seis,eind2} is due to the inclusion of the CHOOZ data which gives
a $\chi^2$/DOF  of about 6/15.

\section{Allowed areas in the three generation parameter space}
\subsection{Constraints on the $\Delta$m$^2_{31}-\tan^2\theta_{13}$
plane}

For the chosen mass spectrum and mixing matrix
the relevant survival probabilities for atmospheric neutrinos
depend on the parameters $\theta_{23}$, $\theta_{13}$ and
$\Delta$m$^2_{32} (\simeq\Delta$m$^2_{31})$ \cite{gg3g} while the CHOOZ
survival probability P$_{\bar{e}\bar{e}}$ depends mainly  on
$\theta_{13}$ and $\Delta$m$^2_{31}$ and very
mildly on $\theta_{12}$ and $\Delta$m$^2_{21}$.
In fig. 1  we plot the allowed domains in the
$\tan^2\theta_{13}-\Delta m^2_{31}$ parameter space the from
analysis of only the CHOOZ data keeping all other parameters free.
We give this plot both with and without taking into account the interference
term. The effect of the interference term is to lift the 
allowed ranges of $\Delta m^2_{31}$.
The shaded area  marked by arrows in this figure
is the allowed range from a combined analysis of 1289 day
atmospheric data and CHOOZ data taken from \cite{fl3g}. At 99\%
C.L. the atmospheric+CHOOZ analysis allows
$\tan^2\theta_{13}\stackrel{<}{\sim} 0.08$ and
1.5$\times$10$^{-3}$eV$^2$
$<\Delta$m$^2_{31}<6.0\times$10$^{-3}$eV$^2$. It also becomes
apparent from this figure that for
$\tan^2\theta_{13}\stackrel{<}{\sim} 0.03$, all values of
$\Delta$m$^2_{31}$ in the range $[$1.5,6.0$]\times$
10$^{-3}$eV$^2$ are allowed at 99\% C.L. where as for
$0.03\stackrel{<}{\sim}\tan^2\theta_{13}\stackrel{<}{\sim}0.075$,
certain values of $\Delta m^2_{13}$ get excluded. A closer
inspection of fig. 1 shows that around $\tan^2\theta_{13} \sim
0.03$ a window in $\Delta m^2_{31}$  is disallowed whereas for
higher values of $\tan^2\theta_{13}$ certain regions of $\Delta
m^2_{13}$ towards higher values of the interval  
$[$1.5,6.0$] \times 10^{-3}$ eV$^2$ 
get disallowed.  The
width of the disallowed range in $\Delta$m$^2_{31}$ depend on
$\tan^2\theta_{13}$. Clearly the $\Delta$m$^2_{31}$ is restricted
more from the atmospheric data while the  more stringent bound on
tan$^2\theta_{13}$ comes from the CHOOZ results.
It is also evident that the region in $\Delta m^2_{31}$ 
which is disallowed in the only CHOOZ 
contour once the interference effects are taken into account is also being 
disallowed by the combined atmospheric and CHOOZ analysis. 
In figs. 2a, 2b and 2c we plot the $\chi^2_{\odot}$,
$\chi^{2}_{CHOOZ}$, and  $\chi^2_{\odot}+ \chi^2_{CHOOZ}$
respectively against $\tan^2\theta_{13}$, keeping $\theta_{12}$,
$\Delta$m$^2_{21}$ and $\Delta$m$^2_{31}$ (in the range
$[$1.5,6.0$]\times$10$^{-3}$\\eV$^2$) free. It is clear from the
three figures that the most stringent bound on $\tan^2\theta_{13}
$($<$ 0.065 at 99\% C.L.) comes from the combined solar and CHOOZ
analysis. 
The pre-SNO bound on $\tan^2\theta_{13}$ that we get 
from the combined solar+CHOOZ
analysis is $\tan^2\theta_{13} \lsim 0.075$.
Thus SNO is seen to tighten the constraint
on the $\theta_{13}$ mixing angle such that the most stringent
upper limit on $\theta_{13}$ is obtained from the solar plus CHOOZ
analysis.

\subsection{Probing the $\Delta$m$^2_{21}-\tan^2\theta_{12}$
parameter space.}

We now attempt to explore the 1-2 parameter space from a combined
solar+CHOOZ analysis, in the light of new results from SNO. The
parameters $\theta_{12}$ and $\Delta$m$^2_{21}$ are mainly
constrained from the solar data. We present in fig. 3 the allowed
areas in the 1-2 plane at 90\%, 95\%, 99\% and 99.73\%  confidence
levels for different sets of combination of $\Delta m^2_{31}$ and
$\tan^2 \theta_{13}$, lying within their respective allowed range
from atmospheric+CHOOZ and solar+CHOOZ analysis. The CHOOZ data
limits the upper allowed range of $\Delta m^2_{12}$ in the LMA
region to $3 \times 10^{-4}$ eV$^2$. In the three flavor scenario
also there is no room for SMA MSW solution at the 3$\sigma$ level
(99.73\% C.L)\footnote{We find that for values of
$\tan^2\theta_{13}>0.25$, one gets allowed areas in the SMA region
at 3$\sigma$ level even after including the SNO data. Beyond this
value of $\tan^2\theta_{13}$ the allowed area in the SMA region
increases and finally for larger values of $\tan^2\theta_{13}$ the
SMA and LMA regions merge with each other. However these large
values of $\tan^2 \theta_{13}$ lie outside the range allowed by
CHOOZ.}. We see from fig. 3 that the allowed regions reduce in
size as we increase $\tan^2\theta_{13}$ for a fixed $\Delta
m^2_{31}$. At the upper limit of the allowed range of
$\Delta$m$^2_{31}$ the LOW solution gets completely disallowed
beyond tan$^2\theta_{13} \sim$ 0.02 while the LMA solution gets
disallowed beyond tan$^2\theta_{13} \sim$ 0.03. At the lower limit
of of $\Delta$m$^2_{31}$ the LMA solution is found to disappear at
99\% C.L. beyond tan$^2\theta_{13} \sim 0.065$, which is the upper
bound of $\tan^2\theta_{13}$   at 99\% C.L., obtained from
solar+CHOOZ analysis. On the other hand for any given
$\tan^2\theta_{13}$ the least allowed area in
tan$^2\theta_{12}-\Delta$m$^2_{21}$ parameter space occurs at
$\Delta$m$^2_{13} \sim$ 4.0$\times$10$^{-3}$ eV$^2$, whereas above
and below this value larger regions of parameter space are
allowed. To illustrate this in fig. 4 we plot the $\chi^2_{\odot}
+ \chi^2_{CHOOZ}$ vs. $\Delta m^2_{31}$ for fixed
$\tan^2\theta_{13}$ allowing the other parameters to vary freely.
The highest value of $\chi^2$ is seen to come for $\Delta m^2_{31}
= 0.004$ eV$^2$ explaining the least allowed area at this value.
The figure also illustrates the occurrence of a disallowed window
in $\Delta m^2_{31}$ around $\tan^2\theta_{13} \sim $ 0.03, as
discussed earlier. Since the solar probabilities are independent
of $\Delta m^2_{13}$ it is clear that the CHOOZ data is
responsible for this feature.
We have plotted these figures taking the interference term in the 
CHOOZ probability into account. However we have explicitly checked 
that the interfernce term in the CHOOZ probability does not have any impact 
on the allowed  
area in the $\Delta m^2_{21}-\tan^2\theta_{12}$ plane. 
There are two reasons for this. The interference term comes multiplied 
with $s_{13}^2$ which is confined to very small values. Also the 
contours that we have plotted are for values of $\Delta m^2_{31}$ 
$> 1.5 \times 10^{-3} $ eV$^2$ as allowed by the combined atmospheric and 
CHOOZ analysis. As is seen from fig. 1 in this region the interference 
term does not have any significant effect. 

\section{Summary, Conclusions and Discussions}
We have performed a three-generation analysis of the solar neutrino and
CHOOZ  data including the recent SNO CC results. 
The mass spectrum considered is one where $\Delta m^2_{21} =
\Delta m^2_{\odot}$ and $\Delta m^2_{31} \approx \Delta m^2_{32} =
\Delta m^2_{atm}$ = $\Delta_{CHOOZ}$. The other parameters are the
three mixing angles $\theta_{13}$, $\theta_{12}$ and
$\theta_{23}$.
 For
the combined solar and CHOOZ analysis the probabilities are
independent of $\theta_{23}$. The solar neutrino probabilities
depend on $\Delta m^2_{21}$, $\theta_{12}$ and $\theta_{13}$. The
CHOOZ probability depends mainly on $\Delta m_{31}^2$ and
$\theta_{13}$ whereas for $\Delta m^2_{12} \gsim 3 \times 10^{-4}$
eV$^2$ it depends  also on  $\Delta m^2_{12}$ and $\theta_{12}$.
The most stringent constraint on the parameter 
$\Delta m^2_{31}$ comes
from the atmospheric neutrino data.
For this we use the updated
values from \cite{gg3g,fl3g}. The combined atmospheric + CHOOZ
analysis gives $\tan^2\theta_{13} \lsim 0.075$ \cite{fl3g,gg3g}.
We keep $\Delta m^2_{31}$ in the range allowed
by the atmospheric neutrino data and determine the allowed values
of $\theta_{13}$ from a combined analysis of solar and CHOOZ data.
The inclusion of the
SNO results puts a more restrictive bound on  $\theta_{13}$
-- $\tan^2\theta_{13} < 0.065$. 
The best-fit comes in the LMA region of the $\Delta
m^2_{21} - \tan^2\theta_{12}$ plane with $\tan^2\theta_{13} = 0.0$
{\it i.e.} at the two generation limit. We present the allowed
region in the $\Delta m^2_{21} - \tan^2\theta_{12}$ parameter
space for various values of $\tan^2\theta_{13}$ and $\Delta
m^2_{31}$ belonging to their respective allowed ranges and
determine the changes in the two-generation allowed region due to
the presence of the  mixing with the third generation. Since very
low values of $\theta_{13}$ are allowed from combined solar and
CHOOZ analysis there is not much change in the two generation
allowed regions. No allowed area is obtained in the SMA region at
3$\sigma$ if one restricts $\tan^2\theta _{13}$ to be $<$ 0.065,
as allowed by combined solar and CHOOZ analysis.

The combination of solar, atmospheric and CHOOZ data allows to fix
the elements of the neutrino mixing matrix. The $U_{e3}$ element
is narrowed down to a small range $\lsim 0.255$ from the
solar+CHOOZ analysis including SNO.
The $\theta_{23}$ mixing angle is $\approx
\pi/4$ from atmospheric data \cite{gg3g,fl3g}. This determines the
mixing matrix elements $U_{\mu 3}$ and $U_{\tau 3}$. The
$\theta_{12}$ mixing angle is limited by the solar data and the
tilt is towards large $\tan^2\theta_{12}$. The mixing matrix at
the best-fit value of solar+CHOOZ analysis is
\\
\begin{equation}
U \simeq {\pmatrix {2\sqrt{\frac{2}{11}}  & \sqrt{\frac{3}{11}} & 0
\cr -\sqrt{\frac{3}{22}} & \frac{2}{\sqrt{11}} &
\frac{1}{\sqrt{2}} \cr \sqrt{\frac{3}{22}} & -\frac{2}{\sqrt{11}}
& \frac{1}{\sqrt{2}} }}
\end{equation}
Thus the best-fit mixing matrix is one where the neutrino pair
with larger mass splitting is maximally mixed whereas the pair
with splitting in the solar neutrino range has large {\it but not}
maximal mixing. It is a challenging task from the point of view of
model building to construct such scenarios \footnote{For a recent
study see  \cite{lavoura}.}.

From the perspective of model building an attractive possibility
is one where both pairs are maximally mixed \cite{bimaximal}. Our
two generation analysis of the solar data showed that for the LMA
MSW region maximal mixing is not allowed at 99.73\% C.L. though
it is allowed for the  LOW \cite{bcgk,seis} solution \footnote{See
however \cite{eind2}.}. However  fig. 4 of this paper shows that
three generation analysis allows $\tan^2\theta_{12}$ = 1.0 with
$\Delta m^2_{21}$ in the LMA region at 99.73\% C.L. for $\Delta
m^2_{31}$ in its lower allowed range $\sim 1.5 \times 10^{-3}$
eV$^2$ and for $\tan^2\theta_{13} \sim $ 0.02. As $\Delta
m^2_{31}$ increases $\theta_{12} = \pi/4$ in the LMA region no
longer remains allowed even at 99.73\% level though it remains
allowed in the LOW-QVO region.
 Further narrowing down of the $\Delta m^2_{12}
-\tan^2\theta_{12}$ parameter space is expected to come from
experiments like KamLand and Borexino  which will be able to
distinguish between the LMA and LOW regions.

\vskip 10pt

S.G. wishes to acknowledge the kind hospitality extended to her by
the theory group of Physical Research Laboratory.

\vskip 10pt

Note added: After submission of our revised manuscript a preprint
\cite{petnew} appeared which finds constraints on $|U_{e3}|^2$ from a similar 
three generation analysis of the CHOOZ data.

\newpage

\begin{table}
\caption{The ratio of the observed solar neutrino rates to the
corresponding BPB00 SSM predictions.}
\begin{tabular}{ccc}
\hline
experiment & $\frac{observed}{BPB00}$ & composition\\
\hline Cl &
0.335 $\pm$ 0.029 & $B$ (75\%), $Be$ (15\%)
\\
Ga & 0.584 $\pm$ 0.039 &$pp$ (55\%), $Be$ (25\%), $B$ (10\%)
\\
SK & 0.459 $\pm$ 0.017  & $B$ (100\%)
\\
SNO(CC) & 0.347 $\pm$ 0.027 & $B$
(100\%) \\
 \hline
\end{tabular}
\end{table}
\begin{figure}
\psfig{figure=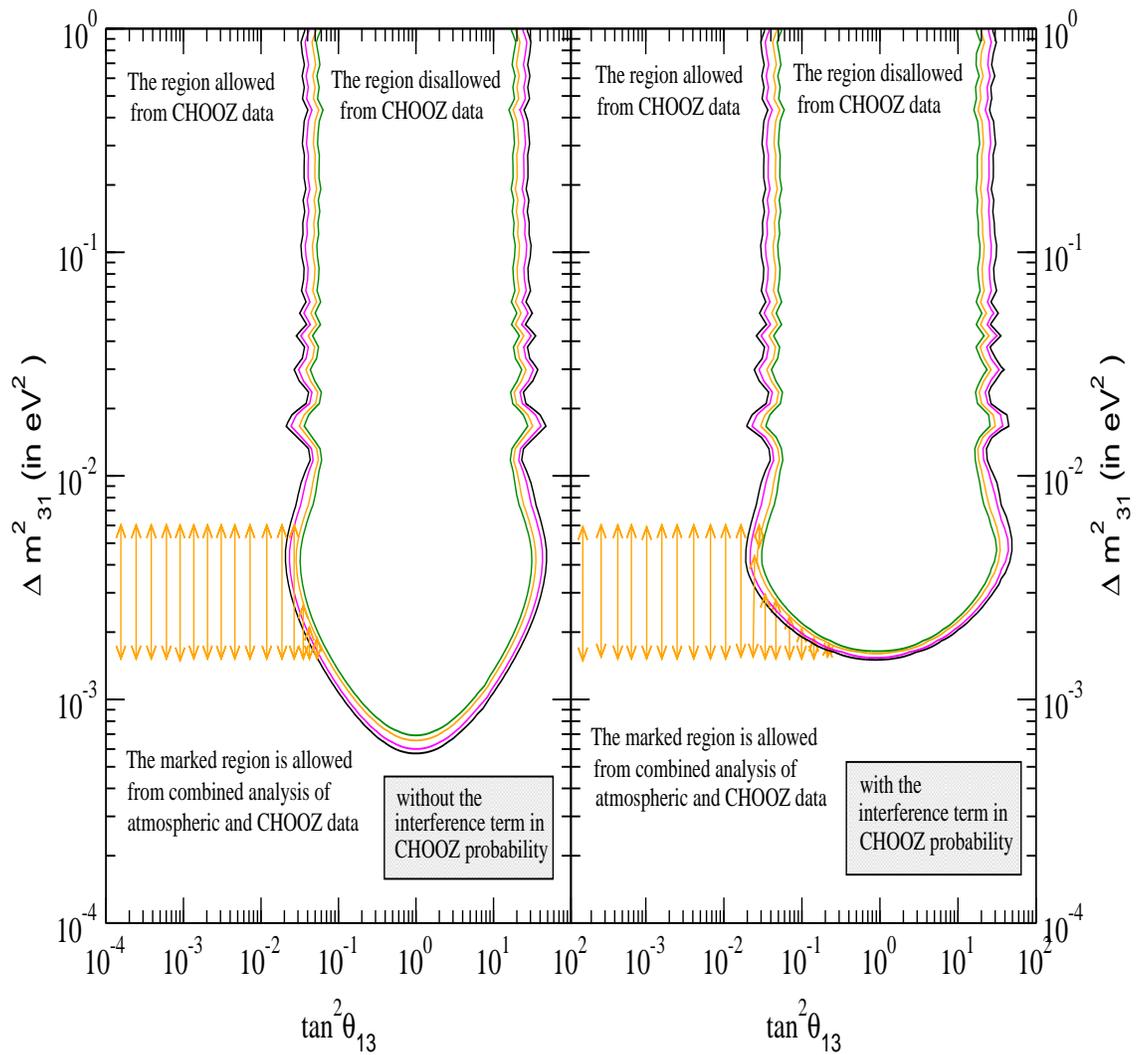,width=16cm,height=17cm,angle=270}
\caption{The allowed areas in (tan$^2\theta_{13}$-$\Delta$m$^2_{31}$) plane
from atmospheric and CHOOZ data.}
\end{figure}
\begin{figure}
\psfig{figure=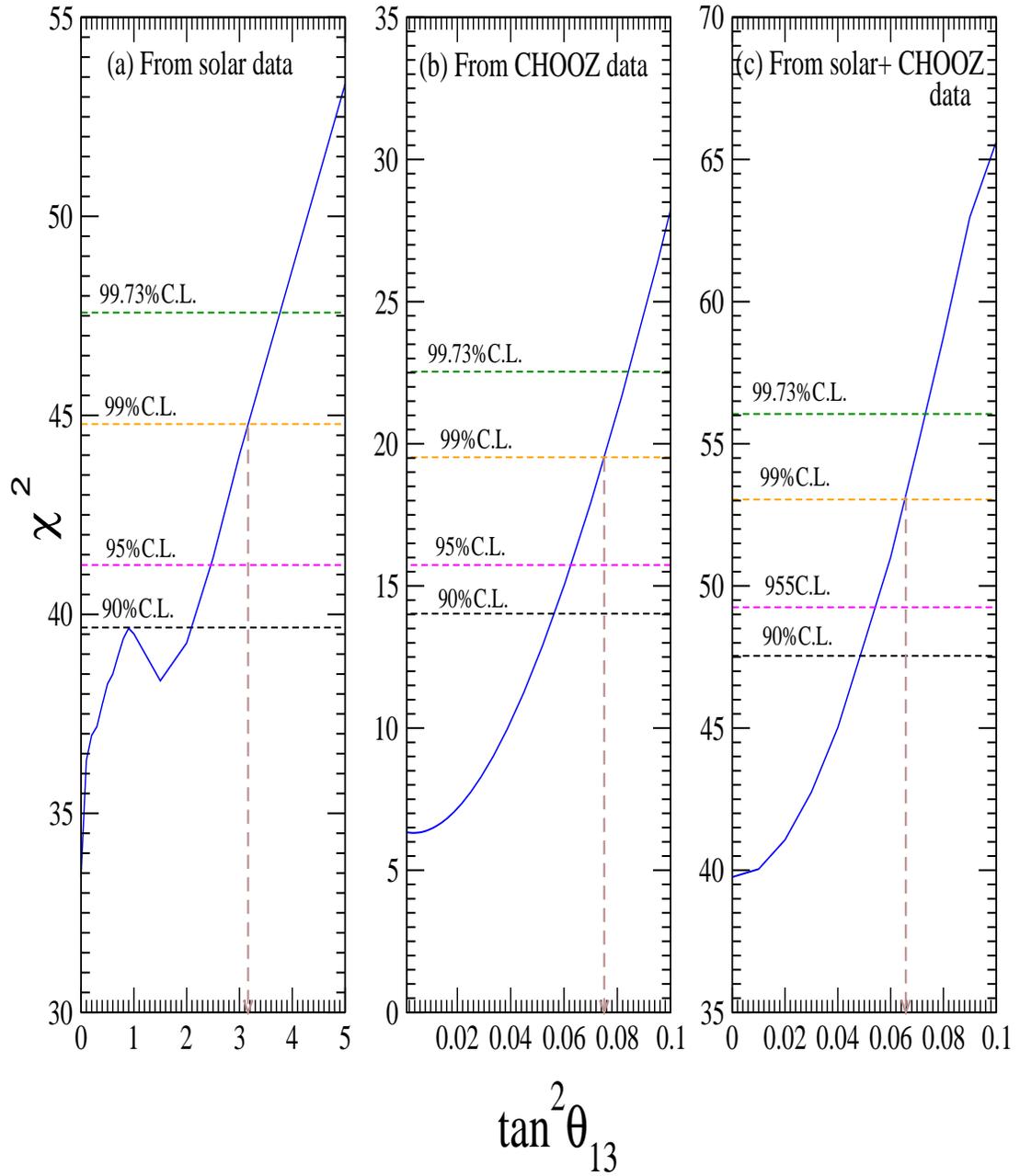,width=18cm,height=18cm,angle=270}
\caption{The plot of $\chi^2$ vs tan$^2\theta_{13}$ from (a) solar  (b) CHOOZ
and (c) solar+CHOOZ data.}
\end{figure}
\begin{figure}
\psfig{figure=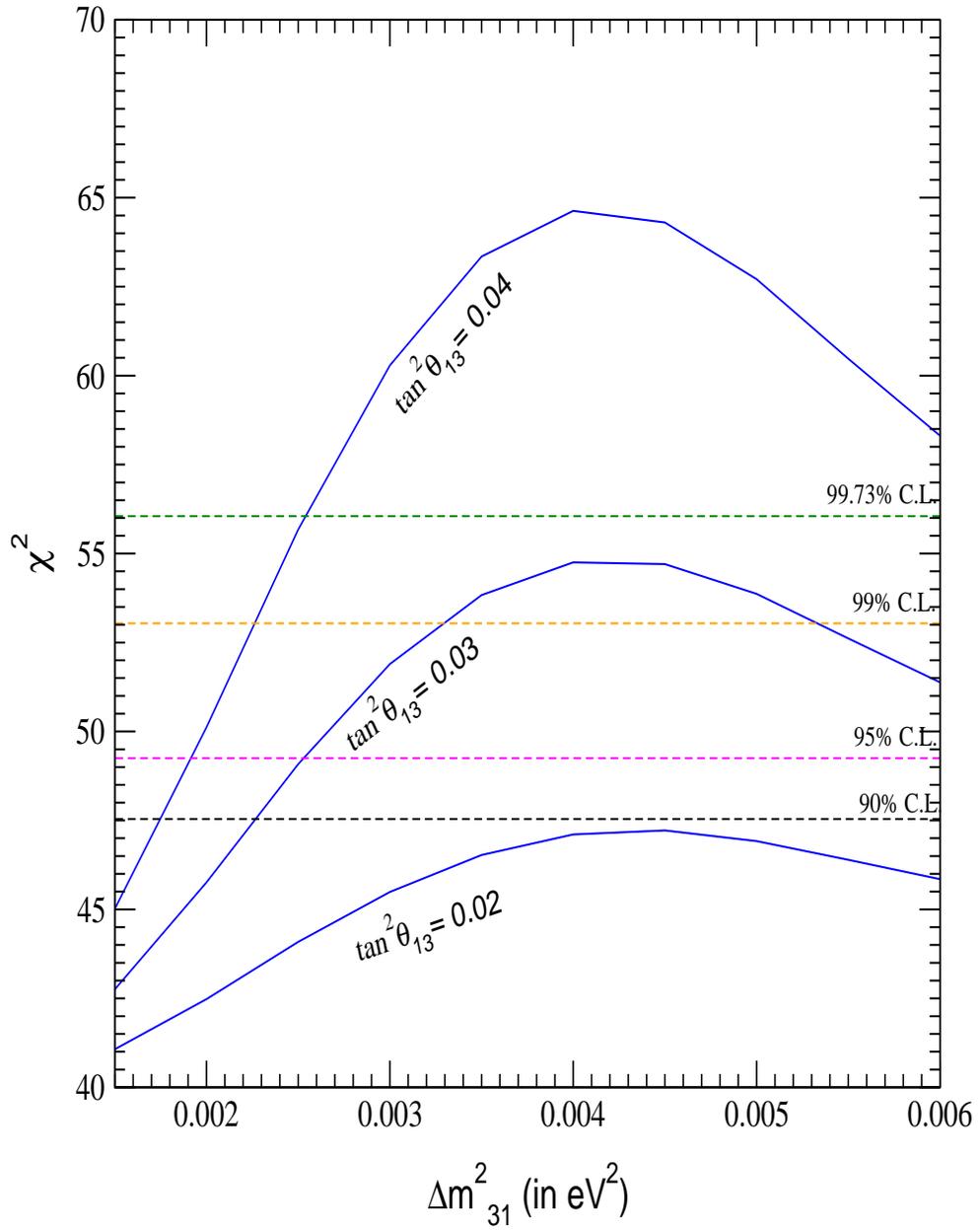,width=18cm,height=18cm,angle=270}
\caption{The plot of $\chi^2$ vs $\Delta$m$^2_{31}$ from solar+CHOOZ data.}
\end{figure}
\begin{figure}
\psfig{figure=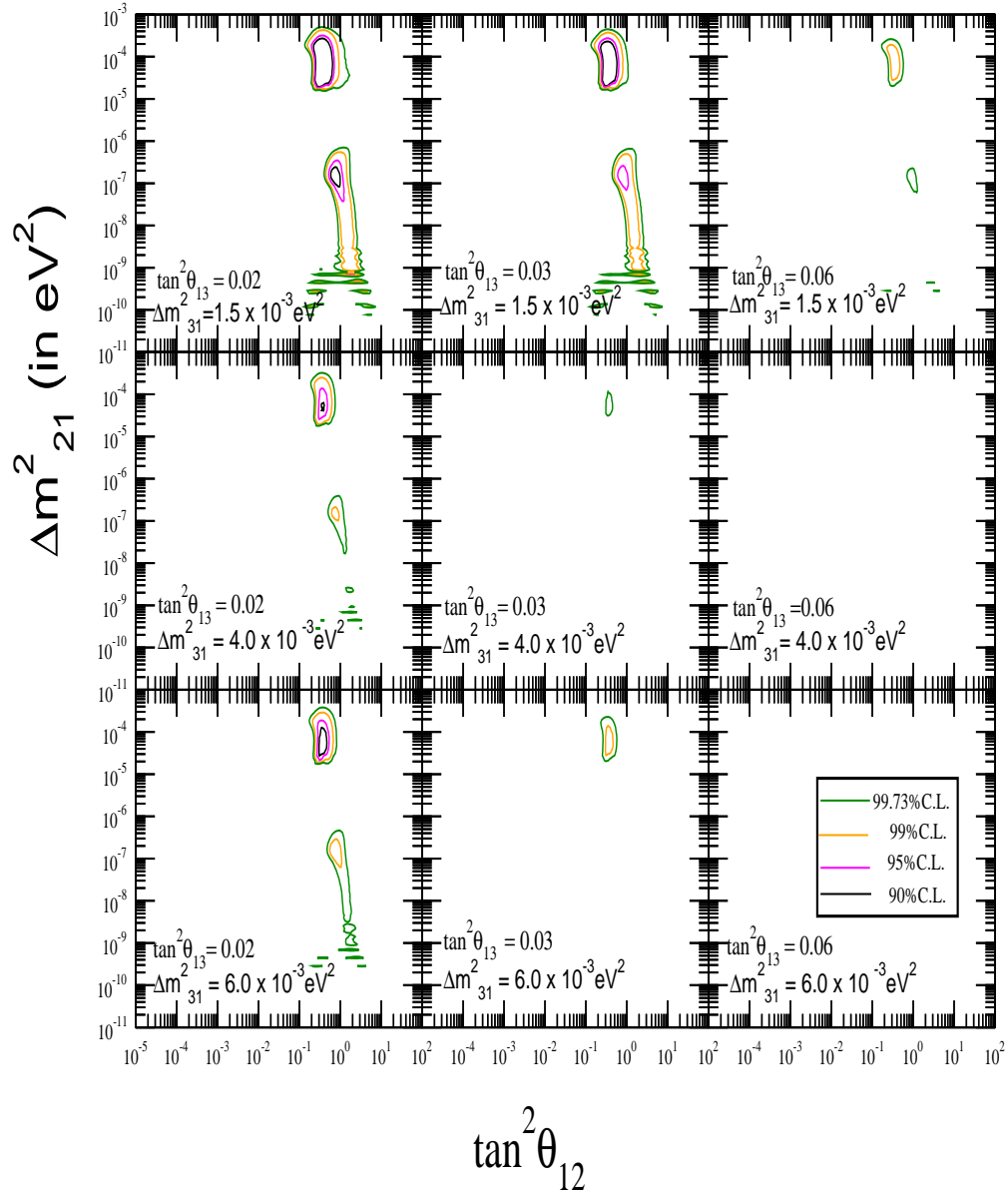,width=18cm,height=18cm,angle=270}
\caption{The allowed areas in (tan$^2\theta_{12}$-$\Delta$m$^2_{21}$) plane
from solar+CHOOZ analysis.}
\end{figure}

\end{document}